# PERFORMANCE ANALYSIS OF MODIFIED ALGORITHM FOR FINDING MULTILEVEL ASSOCIATION RULES

Arpna Shrivastava[1], R. C. Jain[2]

[1]Research Scholar, Barkatullah University, Bhopal
arpna.10878@gmail.com
[2]Director, SATI, Vidisha

## ABSTRACT

*Multilevel association rules explore the concept hierarchy at multiple levels which provides more specific information. Apriori algorithm explores the single level association rules. Many implementations are available of Apriori algorithm. Fast Apriori implementation is modified to develop new algorithm for finding multilevel association rules. In this study the performance of this new algorithm is analyzed in terms of running time in seconds.*

## 1. INTRODUCTION

Association rule learning is a popular and well researched method for discovering interesting relations between variables in large databases. With wide applications of computers and automated data collection tools, massive amounts of transaction data have been collected and stored in databases. Discovery of interesting association relationships among huge amounts of data will help marketing, decision making, and business management. Therefore, mining association rules from large data sets has been a focused topic in recent research into knowledge discovery in databases [1]. Apriori is a classic algorithm for mining frequent item sets and learning association rules of single level [2].

Mining multilevel association rule was first introduced in [3]. Multilevel association rules provide more specific and concrete knowledge. Apriori based algorithm for multiple-level association rules from large database was presented in [5].

Mining association from numeric data using genetic algorithm is explored and the problems faced during the exploration are discussed in [13]. Positive and negative association rules are another aspect of association rule mining. Context based positive and negative spatio-temporal association rule mining algorithm based on Apriori algorithm is discussed in [14]. Association rule generation requires scan of the whole databases which is difficult for very large database. An algorithm for generating Samples from large databases is discussed in [11]. An improved algorithm based on Apriori algorithm to simulate car crash is discussed in [12].

There are many algorithms presented which are based on Apriori algorithm [4,6,7,9]. The efficiency of algorithms is based on their implementation. UML class diagram of Apriori algorithm and its Java implementation is presented in [10]. A fast implementation of Apriori algorithm was presented in [8]. The central data structure used for the implementation was Trie because it outperforms the other data structure i.e. Hash tree.

In this study, the performance of new modified algorithm for finding frequent item sets and mining different level of association rules has been analyzed in terms of running time for different datasets. The different datasets available on Frequent Itemset Mining Dataset Repository (fimi) has been used in this study.

This paper is organized as follows. Problem is defined in section 2. Section 3 discusses the coding of transaction database. Section 4 deals with cleaning of data if required. Modified algorithm is discussed in section 5. Results for running time for different datasets are discussed in section 6. Section 7 deals with conclusion and future scope.

## 2. PROBLEM STATEMENT

The items in any super market are numbered using the barcode. It facilitated the automatic reading of item details using the barcode reader. Barcode for an item can be some logical code or just a sequence number. The transaction database of any super market contains the transaction id and the set of barcodes against each transaction id. The sample transaction table is shown in table 1.

Table 1. Transactional database

| Transaction id | Barcodes |
|---|---|
| 12345 | {121234, 102302, 876546} |
| 12346 | {121212, 102302, 121234} |
| … | ………………………….. |

The item master table contains the details of item against the each barcode. If barcode is just a sequence number then mapping of item details from item master database to transaction database is required to produce some meaningful database. Table 2 shows the sample item master database.

Table 2. Item master database

| Barcode | Category | Brand | Pack | Price (Rs.) |
|---|---|---|---|---|
| 121234 | Bread | Harvest | Normal | 18 |
| 121212 | Milk | Amul | 500ml | 22 |
| …… | …. | …….. | …….. | ………. |

Item master database contains the complete details of items against each barcode. The barcode 121234 represents the item category bread brand harvest, pack normal and price Rs.18/-. This item master is providing three level of concept hierarchy. First level the item category, on the second level the item brand and the third level is pack. By $3^{rd}$ level association rules, the association between normal pack harvest bread with 500ml amul milk will be explored. The different databases available at fimi website have been used for finding the association rules of various levels. The running time of the modified algorithm for finding different level association rules from different databases using fast Apriori implementation is recorded and analysed.

## 3. CODING OF DATA

The algorithm runs of coded database. In this study, the six digits code has been used for every item purchased. The six digits of the code has been divided into three level hierarchy so two digits per level. In this study, maximum hundred categories can be coded. Every category of item can have maximum of hundred brands and every brand for given category can have maximum of hundred packing options. This coding can be flexible in future studies. Using three tables of code and item category, brand and packs, the coding of the database has been done easily.

Sample coding scheme is shown in table 3. Every item category is represented by two digit code. By this approach, maximum of hundred item category can be coded. So after reading the item category the program which is responsible to generate the codes will generate two digits code for every item category.

Table 3. Coding scheme for item categories

| S.No. | Item | Code |
|---|---|---|
| 1 | Milk | 10 |
| 2 | Bread | 11 |
| 3 | Biscuit | 12 |
| 4 | Butter | 13 |
| 5 | Atta | 14 |

Table 4 is showing the sample coding scheme for brand of item category of milk. So every item category can have hundred brands. The program which is responsible to generate the code will put two digits code for the brand name of item category. For example for brand amul of item category milk, the code is 20.

The table 5 is showing the sample code for packing options of items. Generally item comes in various packing options. By this coding scheme, maximum of hundred packing option are available to code the packing options for every brand of item category. For example 200ml pack of brand amul of item category milk 102000.

Table 4. Coding scheme for brands of item category milk

| S.No. | Item | Code |
|---|---|---|
| 1 | Amul | 20 |
| 2 | Mother Dairy | 21 |
| 3 | Sanchi | 22 |
| 4 | Paras | 23 |
| 5 | Jersey | 24 |

Table 5. Coding scheme for packs of brand amul item category milk

| S.No. | Item pack (ml) | Code |
|---|---|---|
| 1 | 200 | 00 |
| 2 | 500 | 01 |
| 3 | 1000 | 02 |
| 4 | 2000 | 03 |

The complete coding scheme of items is shown in table 6. The program will generate six digits code for every item purchased. For example, 102101 is the code for item category milk, item brand mother dairy and packing of 500ml. The results will come in the form of frequent item sets and association rules of 3$^{rd}$ level and decoded easily using these three tables.

Table 6. Coding scheme for milk with brands

| S.No. | Item with brand | Code |
|---|---|---|
| 1 | Amul Milk 200ml | 102000 |
| 2 | Mother dairy milk 500ml | 102101 |
| 3 | Sanchi Milk 200ml | 102200 |
| 4 | Paras Milk 1000ml | 102302 |
| 5 | Jersey Milk 2000ml | 102403 |

Table 7 is displaying the sample transaction table of any super market. Transaction id is assigned against each purchase from the store. For example the first customer purchases the milk of amul brand in 200ml pack, bread of harvest brand in normal pack and ashirvad atta in 2 kg pack.

Table 7. Transaction table

| Tid | Item purchased |
|---|---|
| 1 | {Milk(Amul200ml)),Bread(Harvest(normal)), Atta(Ashirvad(2 kg))} |
| 2 | {Bread(Britania(big pack)), Biscuit(Britania(100gm)),Noodles(Maggi(small))} |
| 3 | {Milk(Amul(500ml)),Bread(Britania(normal)), Biscuit(Parle(100gm))} |
| 4 | {Milk(Mother Dairy(200ml)), Bread(Harvest(normal)),Atta(Ashirvad(2 kg))} |
| 5 | {Milk(Amul(200ml)), Bread(Harvest(normal)), Biscuit(Parle(100gm))} |

The program which is responsible to code the database will generate the data.dat file for the coded database. Each row of this file contains the one row of transaction table. The row number will represent the transaction id and the contents of the row will represent the item purchased against that transaction id. The sample of data.dat file is shown in fig. 1.

```
102000      113001      135002
   113102      124002      146000
102001      113101      124202
102100      113001      135002
102000      113001      124202
```

Figure 1. Data.dat file

This is the input file for the modified Apriori algorithm. The implementation of this modified algorithm will produce the frequent item sets and then the association rules of $3^{rd}$ level. Similarly input file for $2^{nd}$ level association rules is created and used to find frequent item sets and produce $2^{nd}$ level association rules.

## 4. CLEANING OF DATA

Cleaning of data is required for the databases which are already available in coded form. Another program has been developed for cleaning the data files. The program takes the .dat file as input and done the cleaning process. It fills the missing digits by copying the digits are available within the code. After the cleaning process, it generates the new .dat file which has all six digit codes.

The program reads the every code from data.dat file and counts the digits of the code. If code is less than 6 digits it makes the code of six digits by adding the missing digits from the code.

The algorithm of data cleaning is given in fig. 2. The algorithm of data cleaning takes the data.dat file as input and n as the number of required digits in the output file. It opens in read mode and creates and opens out.dat file in write mode. It reads the data.dat file and checks for space and new line character. These characters are the separators between two codes. It stores these codes and writes them into out.dat if the count of the code is less than n digits.

This algorithm return out.dat file which have all n digits code into it. It completes our data cleaning process. It is a complete input file so our algorithm for finding frequent item sets and association rules will work properly.

```
Algorithm Data_Cleaning(data.dat, n)
{
Open data.dat file in read mode
Create and open out.dat in write mode
i=0
While( data.dat)
{
read data.dat into x
if x=' ' or x= new line then
        if i=1 then
        for j =1 to n do
        item[i]=item[i-1]
        if i=2 then
        for j =2 to n do
        item[i]=item[i-2]
        for j=0 to n-I do
        write into out.dat
else
write into out.dat
i=i+1
}
}
```

Figure 2. Algorithm of data cleaning

## 5. ALGORITHM

Apriori algorithm is a classic algorithm for finding frequent item sets and single level association rules [4]. A fast implementation of Apriori algorithm is presented using the trie data structure in [8]. Bodon implementation generates frequent item sets and association rules of single level. It does not generate the association rules of second level.

This Bodon implementation has been modified for finding the association rules of second level. To facilitate the process of finding the level of association rules one argument has been added. The necessary modifications are also done to process this new argument. One additional function is added to separate the code of input file. After separating the coded inputs, it calls the function to generate the association rules according to their required level.

## 6. RESULTS

The new modified algorithm for finding frequent item sets and mining multilevel association rules has been rum on these datasets. The results are recorded for level 1, level 2 and level3 for different minimum support. Every reading is recorded after taking the average of three

Computer Science & Engineering: An International Journal (CSEIJ), Vol. 3, No. 4, August 2013observations. Minimum support has been varied from 0.50 to 0.01. Observations show that running time is increasing with levels and running time is increasing with decreasing minimum support. The system used to run the algorithm has Core 2 Duo (1.6 GHz), and one gigabyte of RAM. The operating system is Red Hat Linux. Clock function is used to calculate the running time. New modified algorithm is used for finding the second and third level association rules and fast Apriori implementation given by Bodon has been used for finding the first level association rules. The T10ID100K and T40I10D100K datasets were generated BY the IBM Almaden Quest research group. The Kosarak dataset was provided by Ferenc Bodon and contains (anonymized) click-stream data of a hungarian on-line news portal. The Retail dataset was donated by Tom Brijs and contains the (anonymized) retail market basket data from an anonymous Belgian retail store. Data cleaning has been done for finding second and third level association rules.

Table 8 shows the results for T10ID100K dataset.

Table 8: T10ID100K dataset

| Min Support | Running Time (Seconds) | | |
|---|---|---|---|
| | Level 1 | level 2 | level 3 |
| 0.50 | 0.38 | 0.44 | 0.51 |
| 0.05 | 0.76 | 0.93 | 1.04 |
| 0.03 | 0.86 | 1.02 | 1.17 |
| 0.02 | 1.00 | 1.15 | 1.30 |
| 0.01 | 1.28 | 1.45 | 1.59 |

Table 9 shows the results for T40I10D100K dataset.

Table 9: T40I10D100K dataset

| Min Support | Running Time (Seconds) | | |
|---|---|---|---|
| | Level 1 | level 2 | level 3 |
| 0.50 | 1.42 | 1.64 | 1.90 |
| 0.05 | 4.37 | 4.91 | 5.32 |
| 0.03 | 5.02 | 5.50 | 6.00 |
| 0.02 | 5.99 | 6.57 | 7.07 |
| 0.01 | 35.26 | 36.14 | 36.29 |

Table 10 shows the results for Kosarak dataset.

Table 10: Kosarak dataset

| Min Support | Running Time (Seconds) | | |
|---|---|---|---|
| | Level 1 | level 2 | level 3 |
| 0.50 | 0.84 | 0.88 | 1.02 |
| 0.05 | 1.82 | 1.95 | 2.17 |
| 0.03 | 1.86 | 1.95 | 2.19 |
| 0.02 | 1.92 | 2.02 | 2.26 |
| 0.01 | 2.10 | 2.19 | 2.43 |

Table 11 shows the results for Retail dataset.

Table 11: Retail dataset

| Min Support | Running Time (Seconds) | | |
| --- | --- | --- | --- |
| | Level 1 | level 2 | level 3 |
| 0.50 | 0.36 | 0.40 | 0.47 |
| 0.05 | 0.73 | 0.84 | 0.99 |
| 0.03 | 0.74 | 0.86 | 1.00 |
| 0.02 | 0.75 | 0.87 | 1.01 |
| 0.01 | 0.81 | 0.96 | 1.08 |

Results show that the new algorithm is effective as the running time for different levels of association rules is proportionate with single level. Association rules of second level and third level provides the more specific information about the databases. Time required for cleaning the dataset is not included in the recorded running time.

## 7. CONCLUSION AND FUTURE SCOPE

The different datasets has been used in this study. The data cleaning has been done before executing the implementation of algorithm for finding second and third level association rules. The new modified algorithm for finding frequent item sets and mining different level association rules is executed on different datasets and running time is recorded for different level of association rules with varying minimum support. The results are acceptable as running time required to find second and third level association rules are proportionate to time required for finding the first level association rules. The running time can be further improved by improving the data structure for storing the item sets.

## REFERENCES

[1] R. Agrawal, T. Imielinski; A. Swami: Mining Association Rules Between Sets of Items in Large Databases", SIGMOD Conference 1993, pp. 207-216.
[2] R. Agrawal, R. Srikant, "Fast Algorithms for Mining Association Rules", Proceedings of the 20th International Conference on Very Large Data Bases, 1994, pp. 487-499.
[3] J. Han, Y. Fu, "Discovery of Multiple-Level Association Rules from Large Database", Proceeding of the 21st VLDB Conference Zurich, Swizerland, 1995, pp.420-431.
[4] R. Agrawal, H. Mannila, R. Srikant, H. Toivonen, and A. I. Verkamo. Fast discovery of association rules. In Advances in Knowledge Discovery and Data Mining, 1996, pp. 307.328.
[5] J. Han, Y. Fu, "Mining Multiple-Level Association Rules in Large Database", IEEE transactions on knowledge & data engineering in 1999, pp.1-12.
[6] Bing Liu, Wynne Hsu and Yiming Ma, "Mining association rules with multiple minimum supports", ACM SIGKDD International Conference on Knowledge Discovery & Data Mining, 1999, pp.337-341.
[7] F. Berzal, J. C. Cubero, Nicolas Marin, and Jose-Maria Serrano, "TBAR: An efficient method for association rule mining in relational databases", Data and Knowledge Engineering 37, 2001, pp.47-64.
[8] F. Bodon, "Fast Apriori Implementation", Proceedings of the IEEE ICDM Workshop on Frequent Itemset Mining Implementations, 2003.

Computer Science & Engineering: An International Journal (CSEIJ), Vol. 3, No. 4, August 2013

[9] N. Rajkumar, M.R. Kartthik and S.N. Sivanandam, "Fast Algorithm for Mining Multilevel Association Rules", Conference on Convergent Technologies for the Asia-Pacific Region, TENCON, 2003, pp.688-692.

[10] Y. Li, "The Java Implementation of Apriori algorithm Based on Agile Design Principles", 3rd IEEE International Conference on Computer Science and Information Technology (ICCSIT), 2010, pp. 329 – 331.

[11] B. Chandra, S. Bhaskar, "A new approach for generating efficient sample from market basket data", Expert Systems with Applications (38), Elsevier, 2011, pp. 1321–1325.

[12] L. Xiang, "Simulation System of Car Crash Test in C-NCAP Analysis Based on an Improved Apriori Algorithm", International Conference on Solid State Devices and Materials Science, Physics Procedia (25), Elsevier, 2012, pp. 2066 – 2071.

[13] B. Minaei-Bidgoli, R. Barmaki, M. Nasiri, "Mining numerical association rules via multi-objective genetic algorithms", Information Sciences (233), Elsevier, 2013, pp.15–24.

[14] M. Shaheen, M. Shahbaz, A. Guergachi, "Context based positive and negative spatio-temporal association rule mining", Knowledge-Based Systems (37), Elsevier, 2013, pp. 261–273.